\documentclass[aps,prb,preprint,longbibliography]{revtex4-1}

\usepackage{graphicx}
\usepackage{hyperref}

\begin{document}

\newcommand{\figA}{
\begin{figure}[t]
\includegraphics[width=\columnwidth]{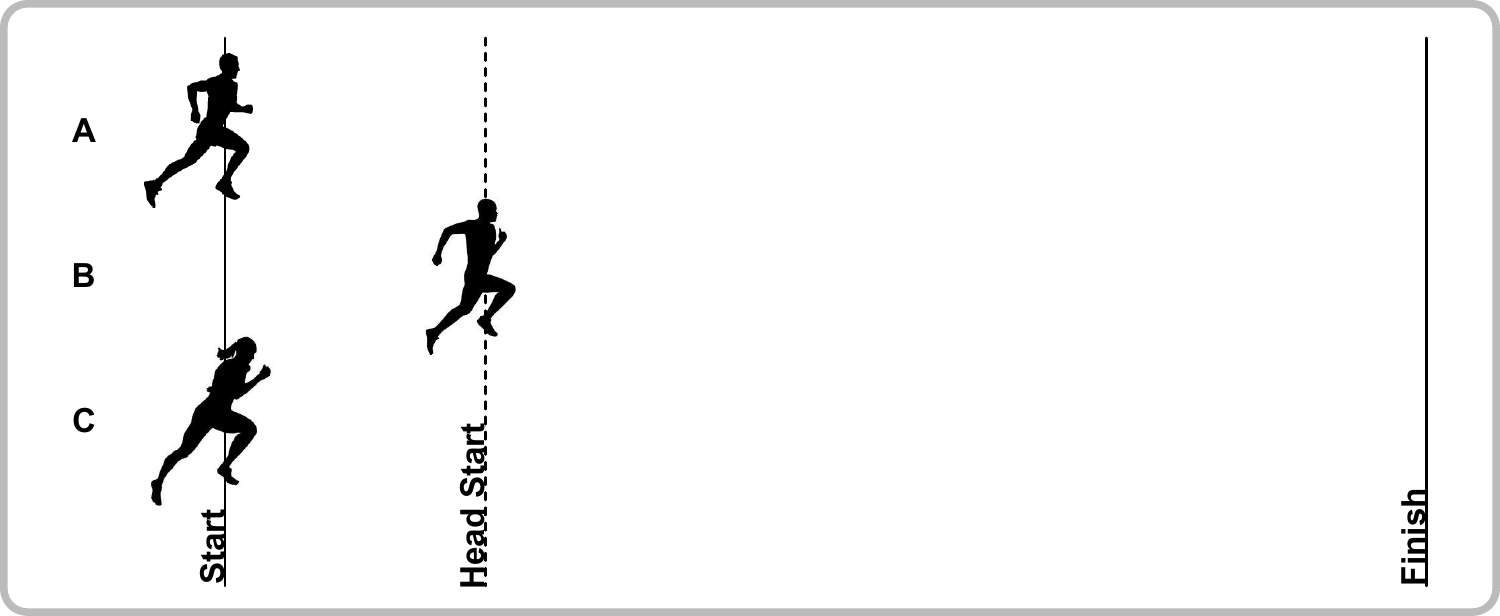}
\caption{\label{fig01} Image used for two different PAR problems. For each PAR problem, the caption of this figure read: ``Three runners are competing in the 100 meter dash. Runner A is faster than Runners B and C. Runner B is new to the sport, and the other runners have agreed to give him a head start in the race. Thus, Runner B starts the race at the head start line rather than the start line. The first runner to pass the finish line wins the race. In previous races, Runner A finished the 100 meter dash in about 10--11 seconds. Runner B, on the other hand, has generally taken 12--13 seconds to run this same distance."}
\end{figure}
}

\newcommand{\figB}{
\begin{figure}[t]
\includegraphics[width=\columnwidth]{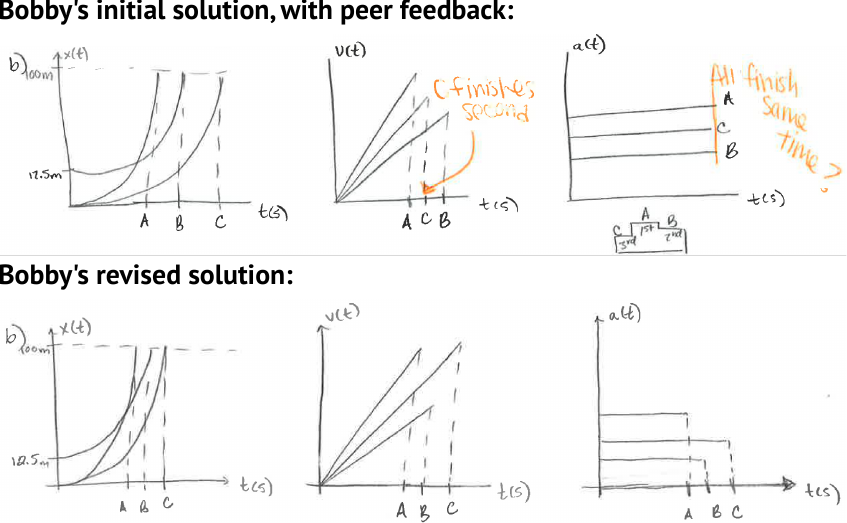}
\caption{\label{fig02} Example of student work before (top) and after (bottom) receiving peer feedback (using a UA model). Bobbie's velocity and acceleration graphs were initially inconsistent with her position graph. After receiving feedback from a peer (orange marks), Bobbie revised her graphs.}
\end{figure}
}

\title{Using Peer Feedback to Promote Reflection on Open-Ended Problems}
\author{Daniel L. Reinholz}
\email{daniel.reinholz@colorado.edu}
\affiliation{Center for STEM Learning, University of Colorado Boulder,
Boulder, CO 80309, USA}

\author{Dimitri R. Dounas-Frazer}
\email{dimitri.dounasfrazer@colorado.edu}
\affiliation{Department of Physics, University of Colorado Boulder,
Boulder, CO 80309, USA}

\date{\today}

%
%

\maketitle

\section{Introduction and background}
This paper describes a new approach for learning from homework, called Peer-Assisted Reflection (PAR).\cite{Reinholz2015c} PAR involves students using peer feedback to improve their work on open-ended homework problems. Collaborating with peers and revising one's work based on the feedback of others are important aspects of doing and learning physics.\cite{DBER2012} While notable exceptions exist,\cite{Nainabasti2015, epstein_immediate_2010, Mason2010AJP, Etkina2006, pierce_applying_2003} homework and exams are generally individual activities that do not support collaboration and refinement, which misses important opportunities to use assessment for learning.\cite{black_developing_2009} In contrast, PAR provides students with a structure to iteratively engage with challenging, open-ended problems and solicit the input of their peers to improve their work. 

Specifically, students are required to: (1)~complete a homework problem outside of class, (2)~reflect on their solution attempt by indicating their level of confidence in their solution and optionally identifying topics to discuss with their peers, (3)~trade their initial attempt with a peer and exchange feedback during class, and (4)~revise their work outside of class to create a final solution. PAR activities engage students both in and out of the classroom, with the peer feedback portion of the cycle typically taking around 10 minutes of class time. All other activities are completed outside of class. At the end of each week-long PAR cycle, students are required to submit a packet containing their initial attempt, reflection, peer feedback, and revised work. 

PAR was designed using principles of formative assessment, which focuses on eliciting information during the learning process and using that information to improve learning activities in real time.\cite{black_developing_2009} Of particular relevance to PAR is peer assessment,\cite{Reinholz2015c} a type of formative assessment where students---rather than teachers---take responsibility for assessing and improving the learning process. Peer assessment is a powerful tool for learning because it helps students learn from their peers and develop the ability to assess their own work.\cite{sadler_formative_1989} The PAR cycle provides a structure for peer assessment, supporting students to provide feedback to each other and revise their solutions based on feedback from their peers. Engaging with PAR on a regular basis supports students to improve in their communication, collaboration, and persistence.\cite{Reinholz2015a} This is consistent with the Next Generation Science Standards, which emphasize scientific practices like communicating information in addition to content mastery and conceptual understanding.\cite{achieve_inc._next_2013}

PAR was initially developed in the context of introductory college calculus,\cite{Reinholz2015a} but has since been used in a variety of settings in both mathematics and the sciences. In the present work, we focus on how we adapted PAR to support pre-service physics teachers to iteratively improve on their solutions to introductory physics problems. We provide examples of student work to illustrate different types of iterative improvement. We also provide practical guidelines for teachers who wish to implement PAR in their own teaching.

\section{Course context}

We adapted PAR for use in an upper-division physics course for students considering a career teaching physics or other physical sciences. The course employed the Modeling Instruction approach to teaching physics,\cite{Brewe2008} a popular teaching strategy that emphasizes looking at similar physical situations with increasingly sophisticated models (e.g., constant velocity versus constant acceleration). Twelve students were enrolled in the course: five were Physics majors, four were majors in other STEM disciplines, and three were Liberal Studies majors (i.e., future K-8 teachers). One of the authors (DRDF) co-taught the course.

In adapting PAR to this setting, several key features of the original design~\cite{Reinholz2015a} were kept the same: use of open-ended problems, format of worksheets, implementation of peer-to-peer conferences, and whole-class discussions of sample student work. However, our implementation differed from the original design in two ways. First, because our students were future teachers, some out-of-class readings focused on physics pedagogy, instructor feedback, and learning more generally. Second, to leverage the iterative nature of Modeling Instruction, some PAR problems used identical physics scenarios that students analyzed using different physics models (as described herein).

Given the iterative nature of PAR, it is a complementary approach to Modeling Instruction. Moreover, Modeling units are initiated with inquiry-based lab activities that are both open-ended and phenomenological in nature. These in-class open-ended activities support the out-of-class PAR homework assignments. PAR focuses on iterative refinements on the scale of a single homework problem; Modeling complements this approach by emphasizing longer-term revisions, as students iteratively develop more sophisticated models as the course progresses. While PAR and Modeling are synergistic approaches, both of them can also be used effectively in isolation. 



\section{Examples of student work}

To illustrate the synergy between PAR and Modeling, we provide student responses to two similar PAR problems. Both PAR problems involve three runners racing in a 100~m dash (Fig.~\ref{fig01}). The first PAR problem was assigned when students were working with Constant Velocity (CV) models. This problem asked students to: compute the speeds of Runners A and B; define a ``fair" head start for Runner~B; determine which runners finished in first, second, and third place; and, plot each runner's position as a function of time.

Later in the quarter, students revisited the 100~m dash scenario in a follow-up PAR problem that was assigned when students were working with Uniform Acceleration (UA) models. In class, the instructors mentioned that races typically require competitors to start from rest, and so CV models fail to describe this scenario completely. In the second PAR problem, students were asked to: determine a ``fair" head start for Runner B using a UA model; plot each runners' position, velocity, and acceleration as functions of time; compare the initial, average, and final speeds of the accelerating runners to the speeds computed in the first PAR problem; and, describe the tradeoffs of using CV or UA models to describe this scenario.

In both cases, students were required to iterate on their homework solutions using the PAR approach. Thus, students were able to iterate on the 100~m dash problem at two timescales: PAR provided opportunities for revision on small timescales while Modeling allowed for a larger arc of iteration and revision in the course.

The open-ended nature of the problem gave rise to a breadth of responses. For example, each student articulated a different definition of ``fair," resulting in different values for the head start. For the CV version of the problem, there were nine distinct recommended distances for the dead start, ranging from 8--17~m. In addition, one student argued that any head start would be inherently unfair, and that Runner B would simply not be able to win the race.

To illustrate how our students responded to these activities, we present responses from three students: Alex, Bobbie, and Corey. These students were chosen because their work demonstrates a diversity of responses to our open-ended prompts. Using Alex's and Bobbie's work as examples, we show how peer feedback can help students clarify their reasoning as well as their position, velocity, and acceleration graphs. Corey's work illustrates how PAR provides opportunities for students to iterate on their reasoning over the course of successive Modeling units. For each student, we show how responses did or did not change when taking into account the runners' nonzero acceleration. 

\subsection{Example 1: Alex}
In his initial solution to the CV PAR problem, Alex wrote the following:
\begin{quote}
\emph{A fair head start should be 17 meter ahead of the start line. That way according to Runner A's and B's rates they will finish at the same time.}
\end{quote}
Here Alex defined a fair head start such that the two runners A and B tie the race. Alex assumed both runners run their slowest, though he did not articulate this assumption. Moreover, his explanation did not explicate how his definition was used to compute a head start of 17~m, an omission which caught the attention of his peer reviewer:
\begin{quote}
\emph{I am confused that they will finish at the same time. Show how you got 17 meters.}
\end{quote}
Alex responded to this feedback by elaborating his reasoning in his revised solution:
\begin{quote}
\emph{A fair head start for runner B would be 17 meters ahead of the start line. This is because in ten seconds Runner A can go 100 meters and in 10 seconds Runner B can go 83 meters based on their average times. So, if Runner B is 17 meters ahead of Runner A, they will both get to the finish at the same time.
} 
\end{quote}
By choosing a head start of 17~m, Alex reasons that Runners B will only have to run 83~m and will therefore finish the race at the same time as Runner A, in alignment with Alex's definition of a fair head start. This is an example of a PAR revision cycle: given feedback from a peer reviewer that his reasoning was not clear, Alex provided further elaboration in his solution.

Later in the semester when working on the UA PAR problem, Alex's definition of ``fair" was unchanged:
\begin{quote}
\emph{A fair head start would be a distance advantage for runner B such that A and B arrive at the finish line at the same time. Assuming Runner A and B run their personal best times of 10 sec and 12 sec respectively the head start line should be placed 30.55~m from the start line. This is farther of a gap than I had given runner B on my previous assessment. This result is dramatically different because we are no longer assuming that the runners begin the race at maximum velocity and maintain it through the race. We are taking into consideration the runners' speeding up, and Runner A speeds up faster than runner B.
} 
\end{quote}
Despite using the same definition of ``fair," Alex nevertheless computes a much larger head start using a UA model than when using a CV model (30.55~m compared to 17~m). Alex correctly attributes this change to the fact that Runner A has a larger acceleration than Runner B.

\subsection{Example 2: Bobbie}

Whereas Alex used a tie condition to determine a fair head start for Runner B, Bobbie used proportional reasoning to infer the head start based on the distances in the diagram in Fig~\ref{fig01}. For example, in her solution to the CV PAR problem, Bobbie wrote:
\begin{quote}
\emph{Judging from the diagram, I gave runner B a head start of 20~m in part~(a). After running through the calculations, I think this head start is too generous. A head start of 10--15~m is more appropriate as it would put runner B at second place. This gives B the chance to win or lose.
} 
\end{quote}
After determining that 20~m was too generous, Bobbie modified her suggested head start. Bobbie's reasoning was that a head start should help Runner B, but should not make him win. Unlike for Alex, Bobbie's reasoning resulted in the same head start for Runner B even when using a different kinematic model for the UA PAR problem:
\begin{quote}
\emph{Runner B should have the same head start that I set before $\sim$12.5~m because all of the runners still finish in the same amount of time despite having to accelerate from rest.
} 
\end{quote}
In both the CV and UA cases, the PAR cycle did not result in Bobbie revising her reasoning.

Nevertheless, through the UA PAR problem, Bobbie was prompted to revise her velocity and acceleration graphs, as can be seen in Fig.~\ref{fig02}. In her initial solution to the UA PAR problem, Bobbie's velocity and acceleration graphs were qualitatively inconsistent with her position graph. In addition to marking on the graphs (Fig.~\ref{fig02}), Bobbie's peer reviewer also made the following comment:
\begin{quote}
\emph{in your V vs t and A vs t graphs you show inconsistency in who finishes first, 2nd, \& 3rd}
\end{quote}
In response to this feedback, Bobbie revised her work accordingly, generating position, velocity, and acceleration graphs that were consistent with one another (Fig.~\ref{fig02}).

\subsection{Example 3: Corey}
While Alex and Bobbie used the same reasoning to determine Runner B's head start in both the CV and UA versions of the PAR problem, Corey changed his definition of ``fair," leading him to use very different mathematical approaches to solving the same problem in these different contexts.

In the CV PAR problem, Corey determined that the head start should be only 8~m:
\begin{quote}
\emph{As stated before in one of my assumptions, Runner BÕs head start was intended to reduce his time by about 1 second. Since he can run 8 meters in one second, his head start time was set 8 meters ahead of the usual line.
} 
\end{quote}
Here Corey decided that ``fair" meant shaving a second off of Runner B's average time. However, in the UA PAR problem, Corey embraced a different definition of ``fair":
\begin{quote}
\emph{I'll define ``fair" such that Runner B, at his best (100~m in 12~seconds), would finish the altered race in 10.5~seconds, which is the average or ``typical" time for the other runner (or at least a good estimate thereof). This will encourage all runners to do their best to try to win. 
} 
\end{quote}
In the UA PAR problem, Corey's definition of ``fair" involved comparing Runner B's fastest speed to Runner A's average speed. In this example, PAR and Modeling complement each other quite nicely: Modeling provides opportunities to revisit similar scenarios using increasingly sophisticated models, and the open-ended nature of PAR problems provides opportunities for students to modify their underlying assumptions and hence the mathematical reasoning they bring to bear on the problem.

\subsection{Discussion}
As Alex, Bobbie, and Corey's solutions demonstrate, PAR facilitates reflection on diverse approaches to solving physics problems. Students differed in their reasoning about what it means to have a fair head start, and they also differed in their approaches to computing the head start. These are just three of numerous responses illustrating how the open ended prompt of a fair head start allowed students to provide a variety of solutions. This is especially important in the context of PAR, because when students enter a peer conference with different solutions, it provides them with a productive space for discussing their differences. For example, some students' definition of ``fair" led them to compare the slowest speed of Runner~A to the fastest time of Runner~B while others' definitions resulted in comparing the runners' average speeds. Through peer conferences, PAR provided a space for students to discuss these different choices and the interpretations of minimum, maximum, and average race times.

Moreover, some students modified their responses with the new model while others did not. Importantly, the PAR problems created explicit opportunities for students to revisit their initial models, make revisions, and make comparisons between models. Using open-ended homework problems such as this in the context of Modeling provided space for students to explore the relationship between the assumptions made in a model and how to define an ambiguous concept such as ``fair.'' In general, revision helped students improve the depth and clarity of their responses.

\section{Guidelines for teachers}

Having illustrated the potential of PAR and modeling to engage students with conceptual reasoning, we now provide guidelines for teachers who wish to use these techniques together in their classroom. 

\subsection{Setup and Logistics}

Typically students are expected to complete a single PAR problem each week. This involves completing an initial solution, bringing that solution to class for peer conferencing, and then turning in a revised solution during a later class period (one or two days later). It is important that students turn in their revised solutions at a later date, to avoid the temptation that some students may have of simply ``getting the problem done'' and turning it in at the end of the same class period.

Prior studies of PAR showed that students benefit from engaging with a variety of partners in PAR.~\cite{Reinholz2015a} This means that students should be encouraged to frequently work with new partners. One way to achieve this is by assigning random partners each week. Logistically, this can also be accomplished by having students sit in a different seat each time before engaging with PAR. 

There are a variety of ways that PAR assignments can be graded. In our own work, we have generally assigned completion points for the self-reflection, peer feedback, and revision portions of the activity. This provides accountability to students to actually engage in the PAR process. For grading students' final work, we have used both standard grading and completion grading in different contexts. 

\subsection{Framing}

To make PAR a successful activity for students, framing is crucial. Assessing the work of their peers and providing feedback is generally not an activity that students have much experience with. Accordingly, there can often be resistance insofar that students feel that they do not have the expertise to provide feedback to their peers. One way to help with this is to emphasize the importance of communication: if you do not understand the problem, a well-explained solution should help you understand it. In this way, it is also possible to provide feedback no matter what your understanding is. A teacher can also emphasize that this is an opportunity to collaborate to come up with solutions, and that in real physics settings, there is no answer book.

It is also important to discuss the types of skills that PAR aims to develop: communication, collaboration, and persistence. A potent framing for the physics contexts is that these are skills that employers highly desire yet physics graduates are often lacking with. It is important than students understand that PAR is aimed both to help them succeed on these given problems and to help them become better at learning physics. 

\subsection{Discussing Sample Work}

Given the novelty of the PAR activity, students can benefit greatly from training on how to give better feedback. In the original studies, this involved weekly class discussions (approximately 10 minutes long) around samples of student work related to the weekly PAR problem. The task for students was to provide helpful feedback to the students who had written the sample solutions, and then the solutions and how to give feedback was discussed as a whole class. This helped students learn to give better feedback, because they had opportunities to practice and discuss with their instructor and peers. As a result of such training, student success in calculus improved greatly, as did the quality of student conversations;~\cite{Reinholz2015b} rather than focusing primarily on getting the ``right answer,'' students focused more on communication and underlying reasoning.

In the context of our physics work, we constructed a variety of activities in which students looked at a variety of solutions from their peers. For instance, with the runners problem, we compiled all of the different handicaps that students suggested and on what basis they determined the handicap. This provided students with chances to see how many different ways one could reason about the problem, and also to learn from the work of their peers. 

\subsection{Discussing Student Feedback}

In addition to training students, it is also very helpful to allocate class time for talking about the types of feedback that students actually give to one another. First and foremost, this communicates to students that the teacher does care about what they are writing, and is willing to devote class time to it. Also, it provides opportunities to speak with students about what feedback they are actually giving and how it can be improved.

\section{Conclusion}

PAR and Modeling are both approaches to teaching physics that support iteration and refinement of one's work. As this paper highlights, these approaches are complementary and can be used to effectively help students iteratively improve their solutions to open-ended physics problems. As students engage with these approaches, they not only learn physics, they learn how to better learn physics.

\acknowledgments The authors acknowledge Chance Hoellwarth and Christine Lindstr{\o}m for contributions to the design of physics PAR activities and feedback on drafts of the paper, respectively. These materials were designed and implemented in the Cal Poly San Luis Obispo Physics Department. DLR and DRDF are supported by the AAU Undergraduate STEM Education Initiative and NSF grant DUE-1323101, respectively. DLR and DRDF contributed equally to this work.


%

\newpage

\figA

\figB

\end{document}